\RequirePackage{ifpdf}
\ifpdf
\else
\PassOptionsToPackage{dvipdfmx}{graphicx}
\PassOptionsToPackage{dvipdfmx}{hyperref}
\fi

\documentclass[a4paper,12pt]{article}
\usepackage{jheppub}

\allowdisplaybreaks[4]
\usepackage{amsmath,bm,ulem}
\usepackage{subcaption}
\usepackage{slashed}
\usepackage{mathrsfs}

\newcommand{\dd}{\mathrm{d}}
\newcommand*{\beqa}{\begin{eqnarray}}
\newcommand*{\eeqa}{\end{eqnarray}}

\title{%
Integrability of Conformal Killing Vectors in the Eisenhart Lift of Scalar-Field FLRW Cosmology
}

\author[a]{Takeshi Chiba,}
\author[b]{and Tsuyoshi Houri}
\affiliation[a]{Department of Physics, College of Humanities and Sciences, Nihon University, Sakurajosui, Tokyo 156-8550, Japan}
\affiliation[b]{National Institute of Technology, Maizuru College, Kyoto 625-8511, Japan}

\emailAdd{chibatak@gmail.com}
\emailAdd{t.houri@maizuru-ct.ac.jp}

\abstract{%
We study the integrability conditions of the conformal Killing equations for the Eisenhart lift of a scalar field in a flat Friedmann-Lema\^\i tre-Robertson-Walker universe. 
The determinant condition of the prolonged conformal Killing equations reduces to a nonlinear second-order differential equation for $h=V'/V$. We solve this equation locally and find two branches. The regular branch reproduces exactly the family of potentials obtained previously, while the singular branch lies on the locus where the determinant equation cannot be written locally in normal form with respect to $h''$ and is incompatible with the full conformal Killing equations. We therefore conclude that the potential obtained in our earlier work is the most general local potential admitting a non-trivial conformal Killing vector in the sector independent of the cyclic Eisenhart coordinate.
}

\begin{document}

\maketitle

\section{Introduction}

The Eisenhart lift, in its original form, geometrizes the trajectories of a classical mechanical system by representing them as geodesics of a higher-dimensional manifold \cite{eisenhart:1928}.  A closely related null-geometric formulation was developed  in terms of Bargmann structures, namely Lorentzian manifolds equipped with a covariantly constant null vector, in connection
with Newton--Cartan theory \cite{Duval:1984cj}.  This  point of view was further developed, where it was shown that non-relativistic many-body dynamics can be encoded in null geodesics of a
higher-dimensional spacetime and its conformal
structures and applications to celestial mechanics were studied \cite{Duval:1990hj}.\footnote{In this sense, 
the construction is sometimes referred to as the Eisenhart--Duval lift
or as a non-relativistic Kaluza--Klein/Bargmann structure.
In this paper, we will consistently  refer to it as the Eisenhart lift,
in recognition of Eisenhart's pioneering work.  }

The relation between the Eisenhart lift, Bargmann 
structure and dynamical symmetries has subsequently been developed in several directions.
In particular, Ref.~\cite{Cariglia:2016oft} 
analyzed time-dependent systems, their projective and conformal symmetries, and 
their quantization from the viewpoint of the Eisenhart 
lift to Bargmann spacetime.  
Cosmological extensions of the
Eisenhart lift was also studied by \cite{Cariglia:2018mos}. 
These works provide a geometric background for the use of lifted metrics and null
geodesic methods in cosmological systems.

The framework of the Eisenhart lift has also been used more widely in
cosmological and minisuperspace settings.  A cosmological extension of the
Eisenhart lifted metric was constructed in Ref.~\cite{Cariglia:2018mos},
where the scale factor and the energy-momentum tensor were
incorporated into the lifted geometry and the Friedmann equations were
derived within the Eisenhart framework.  On the quantum
minisuperspace side, the lift has been used to write covariant and
Dirac-type equations in an extended minisuperspace \cite{Kan:2021edl}, and
later to formulate a third quantization for scalar and spinor wave functions
of the Universe \cite{Kan:2022tq}.  A related spinorial
Wheeler--DeWitt construction has also been discussed for black-hole
interiors \cite{Kan:2023wdw}.  These papers show that the lifted picture is
useful beyond the classical geodesic problem.

In Ref.~\cite{Chiba:2024auh}, hidden symmetries of power-law inflation were studied by using the Eisenhart lift. In Ref.~\cite{Chiba:2024iia}, this method was extended to a homogeneous scalar field with a general potential in a flat Friedmann--Lema\^\i tre--Robertson--Walker (FLRW) universe. There, it was shown that a non-trivial conformal Killing vector (CKV) exists for the potential
\beqa
V(\phi)=V_0\left(\cos\beta \, e^{\alpha\phi}+\sin\beta \, e^{-\alpha\phi}\right)^{-2+\frac{\sqrt{6}}{\alpha}}\,.
\label{potential}
\eeqa
For this class of potentials, the lifted system has an extra conserved quantity and is completely integrable \cite{Chiba:2024auh,Chiba:2024iia}.

For the classical flat FLRW scalar-field system, the same lifted field space was further studied in Ref.~\cite{Chiba:2024iia}. That work found extra Killing and conformal Killing vectors, and in a special case a non-trivial rank-two Killing tensor. Earlier studies of scalar-field minisuperspace also examined exponential-type potentials from the viewpoint of exact solvability and Noether symmetries \cite{Garay:1991,deRitis:1990}. More generally, Killing--Yano and conformal Killing--Yano tensors for Eisenhart--Duval lift metrics have been studied in Refs.~\cite{Cariglia:2012,Cariglia:2014}. The present paper aims to study CKVs of the Eisenhart-lifted field space for the flat FLRW scalar-field model and to ask whether the potential family found in Ref.~\cite{Chiba:2024iia} is already the most general local one.

The derivation of Eq.~(\ref{potential}) in Ref.~\cite{Chiba:2024iia} relied on a factorized ansatz for the components of the CKV. Although that ansatz led to an explicit solution, it did not, by itself, rule out the possibility that further potentials might arise from more general CKVs Our aim here is to settle this point.

Our approach is simple. After briefly reviewing the Eisenhart lift for the system of a scalar field in a flat FLRW universe in Sec. \ref{section2},  in Sec. \ref{section3} we derive the integrability conditions of the reduced CKV equations derived in  Ref.~\cite{Chiba:2024iia} and obtain a necessary differential equation for $h(\phi)=V'(\phi)/V(\phi)$. 
 In Sec.~\ref{section4}, we solve the nonlinear second-order equation obtained from the determinant condition. We first reduce its order by using $h$ as an independent variable and regarding $p=h'$ as a function of \(h\) (Sec.~\ref{section41}). This leads to a first-order equation whose consistency condition splits into two branches (Sec.~\ref{section42}).
The equation has a two-parameter regular branch and a one-parameter singular branch. By returning to the original CKV equations, we show that only the regular branch gives a non-trivial CKV. As a result, in the \(\chi\)-independent sector considered here, Eq.~(\ref{potential}) is the most general local potential admitting a non-trivial CKV.
 Sec. {\ref{section5} is devoted to a summary and discussion.

\section{Eisenhart lift}
\label{section2}

Eisenhart showed that the trajectories of a conservative mechanical system can be reinterpreted as geodesics of a higher-dimensional manifold \cite{eisenhart:1928}. This construction was extended to field theories in Ref.~\cite{Finn:2018cfs}. For a homogeneous scalar field $\phi$ with a potential $V(\phi)$ in a flat FLRW universe with scale factor $a$ whose Lagrangian is given by $-3a\dot a^2+\frac12 a^3\dot\phi^2-V$, the lifted Lagrangian is (see \cite{Finn:2018cfs,Chiba:2024auh,Chiba:2024iia} for details)
\beqa
\mathcal{L}=-3a\dot a^2+\frac12 a^3\dot\phi^2+\frac{1}{4a^3V}\dot\chi^2
\equiv \frac12 G_{AB}\dot\varphi^A\dot\varphi^B\,,
\label{L-FRW}
\eeqa
where $\chi$ is the Eisenhart coordinate, and we have introduced the field-space coordinates $\varphi^A=(a,\phi,\chi)$. Equivalently, the lifted metric reads
\begin{align}
 \dd s^2 = G_{AB}\,\dd\varphi^A\dd\varphi^B = -6a\,\dd a^2 + a^3\,\dd\phi^2 + \frac{1}{2a^3V(\phi)}\,\dd\chi^2\,.
\end{align}
The Hamiltonian constraint of the original cosmological system is mapped to the null condition
\begin{align}
 G_{AB}\dot\varphi^A\dot\varphi^B = 0\,.
\end{align}
Thus, the dynamics of the scale factor and the scalar field is encoded in null geodesics of the lifted field space.

If $\xi_A$ is a CKV,
\begin{align}
 \nabla_{(A}\xi_{B)} = f\,G_{AB}\,,
\end{align}
where $f=\frac13 \nabla_A\xi^A$, then along an affinely parametrized geodesic one has
\begin{align}
 \frac{\dd}{\dd\lambda}\left(\xi_A\dot\varphi^A\right)
 = \dot\varphi^A\dot\varphi^B\nabla_{(A}\xi_{B)}
 = f\,G_{AB}\dot\varphi^A\dot\varphi^B\,.
\end{align}
Hence $\xi_A\dot\varphi^A$ is conserved along null geodesics. In the present system, the cyclic coordinate $\chi$ already provides a trivial Killing vector $\partial_\chi$. Therefore, an additional non-trivial CKV yields an extra conserved quantity, and together with the Hamiltonian constraint the system becomes completely integrable in the sense discussed in Refs.~\cite{Chiba:2024auh,Chiba:2024iia}.

\section{Reduced CKV equations and integrability conditions}
\label{section3}

Following Ref.~\cite{Chiba:2024iia}, we focus on CKVs that are independent of the cyclic coordinate $\chi$ and linearly independent of $\partial_\chi$. 
The remaining independent equations for the covariant components $\xi_a(a,\phi)$ and $\xi_\phi(a,\phi)$ can be written in the system of finite type by introducing
\begin{align}
 \eta(a,\phi) := \partial_a\xi_\phi\,.
\end{align}
The reduced system is then 
\begin{align}
\partial_a \xi_a 
&= -\frac{1}{a}\xi_a+\frac{3h}{a^2}\xi_\phi\,, \label{cka1}\\
\partial_\phi \xi_a
&= \frac{3}{a}\xi_\phi-\eta\,, \label{cka2}\\
\partial_a \xi_\phi
&= \eta\,, \label{cka3}\\
\partial_\phi \xi_\phi
&= \frac{a}{2}\xi_a-\frac{h}{2}\xi_\phi\,, \label{cka4}\\
\partial_a \eta
&= -\frac{3h}{2a}\xi_a+\frac{3(h^2-2h')}{2a^2}\xi_\phi+\frac{2}{a}\eta\,, \label{cka5}\\
\partial_\phi \eta
&= \frac{3h}{2a}\xi_\phi-\frac{h}{2}\eta\,, \label{cka6}
\end{align}
where
\begin{align}
 h(\phi) = \frac{V'(\phi)}{V(\phi)}\,.
\end{align}
 In this system, all first derivatives are written in closed form 
in terms of $\xi_a,\xi_{\phi}$ and $\eta$. 
Throughout, we write $h=h(\phi)$, $h'=\dd h/\dd\phi$ and $h''=\dd^2 h/\dd\phi^2$.

The first non-trivial integrability condition follows from the commutativity of mixed derivatives applied to $\eta$, namely
\begin{align}
 \partial_\phi\partial_a\eta - \partial_a\partial_\phi\eta = 0\,.
\end{align}
Using Eqs.~(\ref{cka1})--(\ref{cka6}), one finds
\begin{align}
 ah'\xi_a - (hh'-h'')\xi_\phi = 0\,.
 \label{ic1}
\end{align}
Differentiating Eq.~(\ref{ic1}) with respect to $a$ and using Eqs.~(\ref{cka1})--(\ref{cka6}) once more, we obtain a second relation,
\begin{align}
 3hh'\xi_\phi - a(hh'-h'')\eta = 0\,.
 \label{ic2}
\end{align}
Differentiating Eq.~(\ref{ic2}) again with respect to $a$ yields
\begin{align}
 ahh''\xi_a - 2h'(hh'-h'')\xi_\phi -2ah''\eta = 0\,.
 \label{ic3}
\end{align}
These three algebraic relations may be assembled as
\begin{align}
M(h,h',h'')
\left(
\begin{array}{c}
\xi_a \\
\xi_\phi \\
\eta
\end{array}
\right)=0\,,
\qquad
M :=
\left(
\begin{array}{ccc}
ah' & -(hh'-h'') & 0 \\
0 & 3hh' & -a(hh'-h'') \\
ahh'' & -2h'(hh'-h'') & -2ah''
\end{array}
\right) .
\end{align}
Therefore, a necessary condition for the existence of a non-trivial solution is the vanishing of the determinant,
\begin{align}
F(h,h',h''):=\det M
=6hh'^2h''+(2h'^2-hh'')(hh'-h'')^2 = 0\,.
\label{intcond}
\end{align}
This is a nonlinear second-order ordinary differential equation for $h(\phi)$.

It is important to stress that Eq.~(\ref{intcond}) is only a necessary condition. Solving it determines the candidate potentials, but each branch must subsequently be checked against the original CKV system.

\section{Factorization and branches}
\label{section4}

\subsection{Reduction of order}
\label{section41}

We first consider non-constant solutions, for which locally
\beqa
h' \neq 0\,.
\label{hneq0}
\eeqa
Then $h$ is locally monotonic and may be used as a new independent variable. Define
\beqa
p(h):= h'(\phi)\,, \qquad u(h):= \frac{\dd p}{\dd h}\,.
\eeqa
By the chain rule,
\begin{align}
 h'' = \frac{\dd p}{\dd\phi} = \frac{\dd p}{\dd h}\frac{\dd h}{\dd\phi}=pu\,.
\end{align}
Substituting this into Eq.~(\ref{intcond}) gives
\begin{align}
 p^3\left[6hu+(2p-hu)(h-u)^2\right]=0\,.
\end{align}
Under the assumption $p=h'\neq 0$, this reduces to the first-order equation
\beqa
6hu + (2p-hu)(h-u)^2 = 0\,.
\label{intcond2}
\eeqa
The case $p=0$ will be discussed separately below.

\subsection{Factorization}
\label{section42}

Introduce a new variable
\beqa
r:= h-u\,.
\label{eq:r}
\eeqa
Then Eq.~(\ref{intcond2}) becomes
\begin{align}
6h(h-r) + \bigl(2p-h(h-r)\bigr)r^2 = 0\,.
\label{4.7}
\end{align}
Solving this equation for $p$ with $r\neq0$, we obtain
\begin{equation}
p = \frac{h(h-r)(r^2-6)}{2r^2}\,.
\label{eq:p}
\end{equation}
The case $r=0$ does not give an additional solution. In this case, Eq.~(\ref{4.7}) implies $h=0$, and then $p=h'=0$. This is the constant-potential case, included in the degenerate solution discussed below.

Now differentiate Eq.~(\ref{eq:p}) with respect to $h$ and impose the consistency condition $\dd p/\dd h=u=h-r$. After a straightforward simplification, one finds
\begin{equation}
\left(h\frac{\dd r}{\dd h}-r\right)(r^3+6r-12h)=0\,.
\label{intcond3}
\end{equation}
This factorisation is the central algebraic step of the analysis. It shows that every non-constant local solution must belong to one of two branches.

\subsection{Branch I: the general branch}

The first factor in Eq.~(\ref{intcond3}) gives
\beqa
h\frac{\dd r}{\dd h}-r = 0\,.
\eeqa
Hence
\beqa
r = mh\,,
\eeqa
where $m$ is an integration constant. Substituting this into Eq.~(\ref{eq:p}) yields
\beqa
p=h' = \frac{1-m}{2}\left(h^2-\frac{6}{m^2}\right)
      = -\frac{\alpha}{\sqrt{6}-2\alpha}h^2+\alpha(\sqrt{6}-2\alpha)\,,
\label{branch1-p}
\eeqa
where we have reparametrised the constant by
\begin{align}
 m = \frac{\sqrt{6}}{\sqrt{6}-2\alpha}
\end{align}
for later convenience. Equation~(\ref{branch1-p}) is a Riccati equation with constant coefficients. Integrating it gives
\beqa
h=\frac{V'}{V}=
(\sqrt{6}-2\alpha)
\frac{\cos\beta \, e^{\alpha\phi}-\sin\beta \, e^{-\alpha\phi}}
     {\cos\beta \, e^{\alpha\phi}+\sin\beta \, e^{-\alpha\phi}}\,,
\label{branch1-h}
\eeqa
where $\beta$ is another integration constant. Integrating once more, we obtain
\beqa
V=V_0\left(\cos\beta \, e^{\alpha\phi}+\sin\beta \, e^{-\alpha\phi}\right)^{-2+\frac{\sqrt{6}}{\alpha}}\,.
\label{branch1-V}
\eeqa
This is exactly the potential (\ref{potential}) found previously in Refs.~\cite{Chiba:2024auh,Chiba:2024iia}. Since the explicit CKV for this potential was already constructed there, Branch~I is an actual solution of the full CKV equations, not merely a solution of the determinant condition.

Branch~I contains two integration constants, $\alpha$ and $\beta$, and therefore provides the general regular  local solution of the second-order equation~(\ref{intcond}). The constant case $p=h'=0$ is included as a degenerate limit. Indeed, $h=C$ gives $V=V_0e^{C\phi}$, and this corresponds to the choices $\beta=0$ or $\beta=\pi/2$ in Eq.~(\ref{branch1-V}); the constant potential, for which $h=0$, is obtained for $C=0$.

\subsection{Branch II: the singular branch}

The second factor in Eq.~(\ref{intcond3}) gives
\beqa
r^3 + 6r - 12h = 0\,.
\eeqa
It is convenient to parametrise this branch by $r$:
\begin{equation}
h = \frac{r^3+6r}{12}\,.
\label{eq:h}
\end{equation}
Substituting Eq.~(\ref{eq:h}) into Eq.~(\ref{eq:p}) yields
\begin{equation}
p=h' = \frac{(r^2-6)^2(r^2+6)}{288}\,.
\label{eq:p2}
\end{equation}
On the other hand, differentiating Eq.~(\ref{eq:h}) with respect to $\phi$ gives
\begin{align}
 h' = \frac{r^2+2}{4}\,r'\,.
\end{align}
Comparing this with Eq.~(\ref{eq:p2}) leads to
\begin{equation}
r' = \frac{(r^2-6)^2(r^2+6)}{72(r^2+2)}\,.
\label{branch2-rprime}
\end{equation}
This equation is separable, and hence Branch~II defines a one-parameter family of local solutions of Eq.~(\ref{intcond}). For the determinant equation alone, it is therefore a singular branch.

\subsection{Compatibility of Branch II with the full CKV system}

We now return to the original CKV equations and show that Branch~II does \emph{not} produce a non-trivial CKV. The key point is that, on Branch~II,
\begin{align}
 hh'-h'' = r h'\,.
\end{align}
Hence Eqs.~(\ref{ic1}) and (\ref{ic2}) become
\begin{align}
 a\xi_a-r\xi_\phi &= 0\,, \label{b2-rel1}\\
 3h\xi_\phi-ar\eta &= 0\,. \label{b2-rel2}
\end{align}
Using Eq.~(\ref{eq:h}), we also have
\begin{align}
 \frac{3h}{r} = \frac{r^2+6}{4}\,.
\end{align}
Therefore,
\begin{align}
 \xi_a = \frac{r}{a}\xi_\phi\,,
 \qquad
 \eta = \partial_a\xi_\phi = \frac{r^2+6}{4a}\xi_\phi\,.
 \label{b2-rel3}
\end{align}
The second relation integrates immediately with respect to $a$:
\begin{align}
 \xi_\phi(a,\phi) = A(\phi)\,a^{\frac{r^2+6}{4}}\,,
 \label{b2-xiphi}
\end{align}
where $A(\phi)$ is an arbitrary function.

Next, Eq.~(\ref{cka4}) implies
\begin{align}
 \partial_\phi\xi_\phi = \frac{a}{2}\xi_a - \frac{h}{2}\xi_\phi
 = \frac{r-h}{2}\xi_\phi\,.
 \label{b2-phi-eq}
\end{align}
Substituting Eq.~(\ref{b2-xiphi}) into Eq.~(\ref{b2-phi-eq}), we find
\begin{align}
 A'(\phi) + \frac{rr'}{2}A(\phi)\ln a
 = \frac{r-h}{2}A(\phi)\,.
\label{b2-logeq}
\end{align}
The left-hand side contains a $\ln a$ term unless $rr'=0$. For a non-constant Branch~II solution, Eq.~(\ref{branch2-rprime}) shows that $r'$ is not identically zero on any open interval. Hence Eq.~(\ref{b2-logeq}) can hold for all $a$ only if
\begin{align}
 A(\phi)=0\,.
\end{align}
It follows from Eqs.~(\ref{b2-xiphi}) and (\ref{b2-rel3}) that
\begin{align}
 \xi_\phi = 0\,,
 \qquad
 \xi_a = 0\,,
 \qquad
 \eta = 0\,.
\end{align}
Thus, Branch~II yields only the trivial solution of the reduced CKV system. The isolated points with $r'=0$ correspond to $r=\pm\sqrt{6}$, for which Eq.~(\ref{eq:p2}) gives $h'=0$; these are precisely the constant solutions already contained in Branch~I.

This also explains why the failure of normal form matters. Let us write Eq.~(\ref{intcond}) as an implicit equation
\begin{align}
 F(h,p,q)=0\,,
 \qquad p=h'\,, \qquad q=h''\,.
\end{align}
At a regular point with $F=0$ and $F_q\neq 0$, the implicit-function theorem allows one to solve locally for
\begin{align}
 q = Q(h,p)\,.
\end{align}
Then the determinant condition behaves like an ordinary second-order equation, and its local solutions have the expected two free constants. This is what happens on Branch~I, and that branch does lift to genuine solutions of the full CKV system.

On Branch~II, however, one finds
\begin{align}
F_q = -h^3p^2 + 4h^2pq - 4hp^3 + 6hp^2 - 3hq^2 + 4p^2q = 0
\end{align}
identically after substituting $q=(h-r)p$. The determinant equation is therefore singular there, and the elimination that produced $F=0$ is no longer regular. For this reason, the reduced equation may admit branches that do not come from the full CKV system. Branch~II is exactly such a branch. Once we substitute it back into the CKV equations, $\xi_\phi$ carries a $\phi$-dependent power of $a$, and Eq.~(\ref{b2-logeq}) then produces a $\ln a$ term that cannot be removed unless the CKV vanishes. So Branch~II solves the determinant condition, but not the original symmetry equations.

\section{Summary \& Discussion}
\label{section5}

In this paper we have studied the integrability conditions of the reduced conformal Killing equations for the Eisenhart-lifted scalar-field system in a flat FLRW universe. We have shown that the determinant of the prolonged system gives the nonlinear second-order equation~(\ref{intcond}) for $h=V'/V$.

We have solved this equation locally. The two-parameter regular branch has reproduced exactly the potential family
\begin{equation}
V(\phi)=V_0\left(\cos\beta \, e^{\alpha\phi}+\sin\beta \, e^{-\alpha\phi}\right)^{-2+\frac{\sqrt{6}}{\alpha}}\,,
\tag{\ref{potential}}
\end{equation}
which was found earlier in Refs.~\cite{Chiba:2024auh,Chiba:2024iia}. We have also shown that the remaining one-parameter branch is singular: the determinant equation cannot be written locally in normal form for $h''$ there, and when we substitute that branch back into the full CKV system, only the trivial CKV remains. Thus, the reduced determinant condition overcounts solutions on the singular locus.

These results have fixed the local classification of potentials in the $\chi$-independent sector. The family~(\ref{potential}) has been shown to be the most general local potential that admits a non-trivial CKV and hence the extra first integral discussed in Refs.~\cite{Chiba:2024auh,Chiba:2024iia}. Our analysis has also illustrated a general point. After one eliminates the symmetry variables and derives a reduced integrability condition, any singular branch of the reduced equation must be checked against the original symmetry equations. 
 It should be emphasized that the classification obtained in this paper is local in $\phi$ and restricted to the sector of CKVs independent of the cyclic Eisenhart coordinate $\chi$. On any connected interval where $V\neq0$ and the reduced CKV system is regular, our analysis shows that a non-trivial CKV can arise only from the regular Branch~I family. A global classification would require additional analysis of the regularity of the potential and the CKV components on the chosen domain. In particular, for the Branch~I potential, possible zeros of
$\cos\beta e^{\alpha\phi}+\sin\beta e^{-\alpha\phi}$
may obstruct a global real and smooth definition of $V$, depending on the value of the exponent $-2+\sqrt{6}/\alpha$. The treatment of such global regularity is beyond the scope of the present work. 

Several extensions remain open.  
The first direction is the non-flat FLRW case. When the spatial curvature is non-zero, the minisuperspace potential picks up an extra term proportional to $a$. The lifted metric then depends on $a$ as well as $\phi$, and the simple reduction to a single function $h(\phi)=V'/V$ is lost. The CKV system is therefore likely to be more complicated than in the flat case. Even so, the same lifted framework has already proved useful in related minisuperspace work \cite{Kan:2021edl,Kan:2022tq}, so this extension looks accessible. It would be interesting to see whether positive or negative spatial curvature only deforms the flat-space family found here, or whether it allows new CKV branches.
The second extension is the multi-field case of Ref.~\cite{Chiba:2024iia}, where one may ask whether an equally sharp classification is possible.

Finally, it would be worth studying other homogeneous cosmologies, higher-rank hidden symmetries, and related quantum problems. The existence of a non-trivial Killing tensor in a special sector of the flat model suggests that a systematic search for Killing tensors, and perhaps Killing--Yano-type objects in related lifted geometries, may sharpen the integrability picture further.

\section*{Acknowledgments}
We would like to thank P. Horvathy for useful comments. 
This work is supported by JSPS Grant-in-Aid for 
Scientific Research Number 22K03640 (TC) and in part 
by Nihon University.

\end{document}